\documentclass[prl, reprint, superscriptaddress]{revtex4-2}

\usepackage[T1]{fontenc}
\usepackage[english]{babel}
\usepackage{graphicx}
\usepackage{amsmath}
\usepackage[utf8]{inputenc}
\usepackage{xcolor}
\usepackage{physics}
\usepackage{overpic,color}
\usepackage{hyperref}
\usepackage{dsfont}

\newcommand{\txtsys}{\text{S}}
\newcommand{\txtcl}{\text{C}}

\newcommand{\numcl}{N_{\txtcl}}
\newcommand{\numsys}{N_{\txtsys}}

\newcommand{\hamiltonianglobal}{\hat{\mathsf{H}}}
\newcommand{\hamiltoniansystem}{\hat{H}}
\newcommand{\hamiltonianclock}{\hat{H}_{\txtcl}}
\newcommand{\interactionfunction}{V}
\newcommand{\interaction}{\hat{\mathsf{\interactionfunction}}}
\newcommand{\interactionclockpart}{\hat{G}}
\newcommand{\interactionsyspart}{\hat{S}}

\newcommand{\potentialsystem}{\hat{V}}
\newcommand{\potentialsystemwanted}{\widehat{W}}

\newcommand{\ident}{\hat{\mathds{1}}}
\newcommand{\identsystem}{\ident}
\newcommand{\identclock}{\ident_{\txtcl}}

\newcommand{\normalization}{\mathcal{N}}

\newcommand{\stateglobal}{\Psi}
\newcommand{\statesystem}{\varphi}
\newcommand{\statesystemwanted}{\phi}
\newcommand{\stateclock}{\chi}

\newcommand{\ketglobal}[1]{\ket{#1}\!\rangle}
\newcommand{\ketsystem}[1]{\ket{#1}}
\newcommand{\ketclock}[1]{\ket{#1}_{\txtcl}}
\newcommand{\braglobal}[1]{\bra*{\!\langle #1}}
\newcommand{\braketsystem}[2]{\braket{#1}{#2}_{\txtsys}}

\newcommand{\braketclockglobal}[2]{\braket*{#1}{#2}\!\rangle_{\txtcl}}

\newcommand{\melclock}[3]{\mel{#1}{#2}{#3}_{\txtcl}}
\newcommand{\melclockglobal}[3]{\mel*{#1}{#2}{#3}\!\rangle_{\txtcl}}
\newcommand{\melglobal}[3]{\langle\!\langle #1 | #2 | #3 \rangle \!\rangle}

\newcommand{\dyadglobal}[2]{\dyad*{#1\rangle\!}{\!\langle#2}}
\newcommand{\dyadclock}[2]{\dyad{#1}{#2}_{\txtcl}}

\newcommand{\projectorclock}{\hat{P}_{\stateclock}}
\newcommand{\projectorclockcomp}{\hat{Q}_{\stateclock}}

\newcommand{\hilbert}{\mathcal{H}}
\newcommand{\hilbertsystem}{\hilbert}
\newcommand{\hilbertclock}{\hilbert_{\txtcl}}

\newcommand{\dimclock}{D_{\txtcl}}

\newcommand{\errorop}{\hat{\mathsf{E}}}
\newcommand{\fidelity}{F}

\newcommand{\sigx}{\hat{X}}
\newcommand{\sigy}{\hat{Y}}
\newcommand{\sigz}{\hat{Z}}

\newcommand{\sigzcl}{\sigz^{\text{(C)}}}

\newcommand{\variance}{\text{Var}}
\newcommand{\costfunc}{\eta}
\newcommand{\envelope}{g}
\newcommand{\enveloperecon}{h}
\newcommand{\weight}{w}
\newcommand{\genericfunc}{q}

\DeclareMathOperator{\sinc}{sinc}

\newcommand{\unitaryvqa}{\hat{\mathsf{U}}_{\text{VQA}}}
\newcommand{\unitarysampling}{\hat{U}_{\textrm{z}}}
\newcommand{\unitaryclockinit}{\hat{U}_{\textrm{C}}}
\newcommand{\prmvqa}{\boldsymbol{\theta}}

\newcommand{\txtinit}{\text{init}}
\newcommand{\qfttxt}{\text{QFT}}
\newcommand{\qft}{\widehat{\qfttxt}}

\begin{document}

\title{Sampling Continuous Quantum Dynamics from a Single Static State}
\author{Sebastian Gemsheim}
\altaffiliation[Present address: ]{SaxonQ, Emilientra\ss e 15, D-04107 Leipzig, Germany}
\affiliation{Max Planck Institute for the Physics of Complex Systems, N\"othnitzer Stra\ss e 38, D-01187 Dresden, Germany\looseness=-1}
\affiliation{Institute for Photonic Quantum Systems, Warburger Stra\ss e 100, D-33098 Paderborn, Germany\looseness=-1}
\author{Felix Fritzsch}
\email{fritzsch@pks.mpg.de}
\affiliation{Max Planck Institute for the Physics of Complex Systems, N\"othnitzer Stra\ss e 38, D-01187 Dresden, Germany\looseness=-1}

\begin{abstract}
While quantum simulation is one of the most promising applications of modern quantum devices, accessible simulation times are fundamentally limited by finite coherence times due to omnipresent noise.
Based on the ideas of relational dynamics/time and of exchanging time for space resources, we propose an approach to simulating quantum dynamics under general time-dependent Hamiltonians in continuous time, which aims to overcome this limitation by encoding the full dynamics in a single static quantum state.
This is achieved by introducing auxiliary qubits, which play the role of a clock, and by tailoring their dynamics and interaction with the original system.
As opposed to traditional methods, no short-time propagators, or approximations thereof, are required in our framework. 
We outline the preparation of a static global state of system and clock via a variational quantum-classical algorithm as well as the sampling of the system's dynamics by performing projective measurements on the clock.
Finally, we provide an example of our approach in terms of a driven qubit.
\end{abstract}

\maketitle

\begin{figure}[h]
    \centering
    \includegraphics[width=0.5\textwidth]{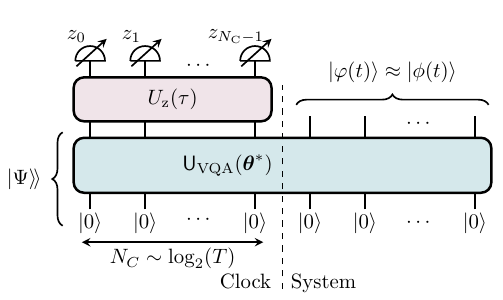}
    \caption{
    Schematic representation of encoding and extracting the relationally evolved state $\ketsystem{\statesystem(t)}$. First, the global state $\ketglobal{\stateglobal}$ is prepared by the optimized unitary $\unitaryvqa(\prmvqa^*)$ acting on both the clock and the system initialized in 
    $\ket{0}^{\otimes \numcl + N_S}$. Subsequently, the clock qubits are transformed into the computational, i.e., the measurement basis via $\hat{U}_z$. Finally, the measurement outcome determines the time $t=t(Z, \tau)$. 
    }
    \label{fig:overview}
\end{figure}

\emph{Introduction}---
Efficiently predicting the dynamics of physical systems from cosmic distances all the way down to atomic systems is fundamental to a plethora of applications.
On the smallest scales, where the dynamical laws are inherently quantum, these include, e.g., simulating molecular dynamics for drug discovery~\cite{Durrant2011} or designing novel materials~\cite{Basov2017}.
While the complexity of simulating quantum dynamics classically grows exponentially with the number of particles, the advent of analog or programmable quantum devices allows for predicting the dynamics of an initial state $\ketsystem{\statesystemwanted(0)}$ under the, in general time-dependent, Schr\"odinger equation 
\begin{equation}
    i \dv{t} \ketsystem{\statesystemwanted(t)}
    = \left( \hamiltoniansystem + \potentialsystemwanted(t) \right) \ketsystem{\statesystemwanted(t)} \,.
    \label{eq:tdse}
\end{equation}
quantum mechanically ($\hbar=1$).
In this so called Hamiltonian simulation~\cite{Feynman1982, Miessen2022}, the system of interest is mapped to the dynamics of the quantum device under the time-independent Hamiltonian $\hamiltoniansystem$ and an external time-dependent potential $\potentialsystemwanted(t)$.
On programmable, gate-based quantum computers, implementing the dynamics induced by the time-dependent Schr\"odinger equation~\eqref{eq:tdse} is typically based on the serial implementation of short-time propagators. 
Consequently, achieving high accuracy by keeping time steps small or simulating late time dynamics requires long coherence times.
As near term devices are noisy, this represents an intrinsic bottleneck for the accessible accuracy and time scales~\cite{Miessen2022}.
\\
These limitations might be circumvented by trading time for space resources, i.e., by reducing the number of gate applications in exchange for controlling a larger number of qubits. 
Such a substitution can be formalized by the so-called "relational approach to time", a powerful idea from foundational research, in which dynamics are mapped onto a static global state $\ketglobal{\stateglobal}$ in an extended Hilbert space.
Roughly speaking, the quantum correlations in the state $\ketglobal{\stateglobal}$ encode all the dynamical information of $\ketsystem{\statesystemwanted(t)}$, with the additional degrees of freedom acting  
as a "clock" (C) for the primary system (S)~\cite{Born1930, Page1983, Wootters1984, Pegg1991, Briggs2001, Giovannetti2015, Marletto2017, Hoehn2021}.
The state $\ketglobal{\stateglobal}$, being static, is formalized by choosing it as an eigenstate of the global Hamiltonian
\begin{align}
    \hamiltonianglobal 
    &= \identclock \otimes \hamiltoniansystem  + \hamiltonianclock \otimes \identsystem + \interaction
    \label{eq:global_hamiltonian}
    \,,
\end{align}
of the composite C+S, 
in which $\hamiltonianclock$ denotes an appropriately chosen Hamiltonian of the clock, whereas $\interaction$ mediates the interaction between the two subsystems.
The system's state is recovered by conditioning the global state on the states of the clock, which thus act as time stamp for the system~\cite{Feynman1986}.
In the presence of an interaction $\interaction$ this procedure yields an effective time-dependent potential $\potentialsystem(t)$ affecting the system's evolution~\cite{Briggs2001, Gemsheim2023} and hence allows for reproducing the dynamics governed by Eq.~\eqref{eq:tdse}. 
An illustrative example, tailored to our intended use case, for a such a composite system hosting the global state $\ketglobal{\stateglobal}$ as well as the conditional system state is shown in Fig.~\ref{fig:overview}.

Approaches, similar to ours, for encoding the full dynamical information in a single global state have previously been proposed. For instance, for time-independent system Hamiltonians, a quantum circuit for preparing a suitable global state has been constructed in Refs.~\cite{Boette2016, Boette2018, Diaz2023}.
Alternatively, even in the presence of a time dependent external potential, the global state can be obtained as a ground state of the so-called \emph{Feynman-Kitaev Hamiltonian}~\cite{Feynman1986, Kitaev2002, McClean2013, Caha2018, Jalowiecki2020, Barison2022, VargasCaldern2023} or \emph{circuit-to-Hamiltonian construction}~\cite{Breuckmann2014}. 
However, all those approaches rely on the implementation of suitable short-time propagators  for the system's dynamics and consequently are limited to discrete times separated by the corresponding time steps $\Delta t$. 
The latter have to be chosen sufficiently small to achieve the desired accuracy $\sim \varepsilon$, which in turn determines the number of necessary qubits in the clock $\numcl \sim\log(T/\Delta t)$ for small time steps. \\

In this letter, we aim to overcome the aforementioned limitations and, based on recent insights~\cite{Gemsheim2023}, present an alternative approach for obtaining relational dynamics in continuous time for arbitrary time-dependent system Hamiltonians in Eq.~\eqref{eq:tdse}.
Contrary to previous methods, we do not presume any specific form for unitary time propagators and focus on energy eigenstates of suitable global Hamiltonians~\eqref{eq:global_hamiltonian} instead. 
The approach outlined below maintains the favorable logarithmic scaling of the number of clock qubits $\numcl \sim \log(T/\varepsilon)$ with $\varepsilon$ the accuracy of the simulation.

Our strategy consists of four general parts, namely 
\emph{(I)} choosing an appropriate $\hamiltonianclock$ and set of clock states $\ketclock{Z}$ which can be uniquely mapped to continuous times $t=t(Z,\tau)=Z \Delta t + \tau$ with an additional offset $\tau$, 
\emph{(II)} tailoring the interaction $\interaction$ such that the effective system potential $\potentialsystem(t)$ obtained within the relational time formalism coincides with $\potentialsystemwanted(t)$, 
\emph{(III)} determining an approximate global energy eigenstate $\ketglobal{\stateglobal}$ by means of a variational quantum algorithm (VQA), and
\emph{(IV)} sampling the relational system dynamics by performing projective measurements in the basis of clock states $\ketclock{Z}$ on the transformed global eigenstate $\unitarysampling(\tau) \ketglobal{\stateglobal} \propto \sum_Z \ketclock{Z}\otimes \ketsystem{\statesystem(t(Z,\tau))}$.
The algorithm and the resulting relational state $\ketsystem{\statesystem(t)}$, which gives an approximation of the true evolved  state $\ketsystem{\statesystemwanted(t)}$ is illustrated in
Fig.~\ref{fig:overview}.
For the remainder of this letter, we provide a brief introduction to relational time before explicating the details of each algorithmic step \emph{(I)}-\emph{(IV)}. 
In order to corroborate our proposition, we present numerical demonstrations and point out opportunities for further research. 


\emph{Relational time}---The simple, but paradigmatic model of two uncoupled qubits~\cite{Marletto2017, Smith2019, Gemsheim2023} allows us to illustrate the key ideas and properties of relational dynamics. 
In this instance, we set $\interaction = 0$ in Eq.~\eqref{eq:global_hamiltonian} for the composite system with $\hamiltoniansystem = -\sigz$ and $\hamiltonianclock = -(1+\epsilon)\sigz_{\txtcl}$ for a small constant $\epsilon \ll 1$. 
Here, $\sigz$ denotes the usual Pauli operator, where $\sigz \ket{0} = \ket{0}$ and $\sigz \ket{1} = -\ket{1}$ in the computational basis.
In order to illustrate the competing factors for obtaining the correct relational dynamics (without time-dependent potentials) in a simple fashion, we focus only on the two central global eigenvalues $\pm \epsilon$ and their respective eigenstates $\ketclock{1} \otimes \ket{0}$ and $\ketclock{0} \otimes \ket{1}$. 
Hence, a general entangled global state in the Hilbert space $\hilbertclock \otimes \hilbertsystem$ comprises the superposition
\begin{equation}
    \ketglobal{\stateglobal} = a \ketclock{1} \otimes \ket{0} + b \ketclock{0} \otimes \ket{1}
\end{equation}
with $\abs{a}^2 + \abs{b}^2 = 1$. The double-angled brackets serve as a reminder of the bipartite global structure. 
While general open system dynamics depends on partial traces representing ignorance of environmental information, the relational approach rests on the idea of conditioning the system state 
\begin{equation}
    \ketsystem{\statesystem} 
    = \frac{1}{\sqrt{\normalization}} \braketclockglobal{\stateclock}{\stateglobal}
    \in \hilbertsystem
    \label{eq:relational_pure_state}
\end{equation}
to a specific clock state $\ketclock{\stateclock} \in \hilbertclock$ through the correlations encoded in $\ketglobal{\stateglobal}$.
Obtaining the conditional state operationally corresponds to a projective measurement of the global state in some basis of the clock.
Since $\braketclockglobal{\cdot}{\cdot}$ represents an integration over clock degrees only, the conditional state lives in the system Hilbert space $\hilbertsystem$ and is normalized through $\normalization = \melglobal{\stateglobal}{\projectorclock}{\stateglobal}$ for $\projectorclock = \dyadclock{\stateclock}{\stateclock}$~\footnote{We neglect explicit identity operators for the sake of notation whenever possible without risk of confusion.}.
As an illustrative example we choose $\ketclock{\stateclock(0)} = (\ketclock{0} + \ketclock{1})/\sqrt{2}=\ketclock{+}$ as initial clock state and aim for an initial system state of the same form, i.e., $\ketsystem{\statesystem(0)} = \ketsystem{+}$.
This can be realized by fixing $a=b=1/\sqrt{2}$.
Transformations of $\ketclock{\stateclock(0)}$ generated by $\hamiltonianclock$ correspond to $\ketclock{\stateclock(t)} = e^{-it\hamiltonianclock} \ketclock{\stateclock(0)}$ and, in order to guarantee the approximate global 
invariance $e^{it\hamiltonianglobal} \ketglobal{\stateglobal} = \ketglobal{\stateglobal}$~\cite{Boette2016, Boette2018, Gemsheim2024}, induce the relational system state
\begin{align}
    \ketsystem{\statesystem(t)} = 
    \frac{1}{\sqrt{2}}\left(e^{it(1+\epsilon)} \ketsystem{0} + e^{-it(1+\epsilon)} \ketsystem{1}\right) \,.
    \label{eq:relational_state_example_intro}
\end{align}
In the resonant case, $\epsilon=0$, $\ketglobal{\stateglobal}$ becomes an energy eigenstate and the above invariance principle, also dubbed \emph{envariance}~\cite{Zurek2003}, is exactly fulfilled, such that both subsystem transformations compensate each other.
Consequently, Eq.~\eqref{eq:relational_state_example_intro} yields the exact dynamics of the true solution $\ketsystem{\statesystemwanted(t)}$ to Eq.~\eqref{eq:tdse} for initial condition $\ketsystem{\statesystemwanted(0)} = \ketsystem{\statesystem(0)}$. 
In the non-resonant case, when $\ketglobal{\stateglobal}$ is only an approximate energy eigenstate, a non-vanishing global energy variance $\variance_{\stateglobal}(\hamiltonianglobal) = \melglobal{\stateglobal}{\hamiltonianglobal^2}{\stateglobal} - \melglobal{\stateglobal}{\hamiltonianglobal}{\stateglobal}^2 = \epsilon^2$ results in
deviations between the relational and the true dynamics.
The latter is conveniently quantified by a decay of fidelity
\begin{align}
    \fidelity(t) = \abs{\braketsystem{\statesystemwanted(t)}{\statesystem(t)}}^2 
    = \cos^2(\epsilon t)
    \approx 1 - \frac{t^2}{2} \variance_{\stateglobal}(\hamiltonianglobal) \, .
\end{align}
Hence, relational dynamics coincides with the exact result for up to times $T$ with
\begin{align}
    T^2 \cdot \variance_{\stateglobal}(\hamiltonianglobal) \ll 1 \,.
    \label{eq:time_energy_uncertainty}
\end{align}
Although derived from a simple model, the final statement holds in full generality and indicates, that accurate relational dynamics requires sufficiently small global energy variance. \\
 
It remains true also in the presence of an interaction $\interaction$ between clock and system, which, similar to Born-Oppenheimer treatments~\cite{Briggs2000, Briggs2001, Schild2018}, additionally induces an effective system potential $\potentialsystem(t)$.
The latter becomes time-dependent as it depends directly on the state of the clock via~\cite{Gemsheim2023}
\begin{align}
    \potentialsystem(t) 
    &= \frac{1}{\normalization(t)} \melclock{\stateclock(t)}{ \Bigl\{ \interaction, \dyadglobal{\stateglobal}{\stateglobal} \Bigr\} }{\stateclock(t)} + \alpha(t) \identsystem
    \label{eq:effective_system_potential}
\end{align}
with 
the anti-commutator $\{ \cdot, \cdot \}$ and an unimportant scalar term $\alpha(t)$. 
This simplifies tremendously in cases of clock states which approximately commute with the interaction 
i.e., $[\projectorclock(t), \interaction] \approx 0$, as the effective system Hamiltonian governing the relational dynamics reduces to
\begin{align}
	\hamiltoniansystem + \melclock{\stateclock(t)}{\interaction}{\stateclock(t)}
	\label{eq:system_hamiltonian_quasi_eigenstate_approximation}
\end{align}
without a dependency on the specific global state~\cite{Gemsheim2023}. 
Substituting Eq.~\eqref{eq:effective_system_potential} by the second term in Eq.~\eqref{eq:system_hamiltonian_quasi_eigenstate_approximation} induces deviations from the exact dynamics, quantified by the (unnormalized) error $\melglobal{\stateglobal}{\errorop}{\stateglobal}$ from
\begin{align}
	\errorop
    &= \frac{1}{T} \int_0^T \dd{t} \projectorclockcomp(t) \interaction \projectorclock(t) \interaction \projectorclockcomp(t) 
    \label{eq:operator_potential_correction}
\end{align}
with $\projectorclockcomp = \identclock - \projectorclock$ (see Supplemental Material).
If the deviations due to the approximation are small, the advantage is two-fold:
An appropriate choice of $\interaction$ and clock state lead to effective potentials $\potentialsystem(t)$, which yield the same dynamics as $\potentialsystemwanted(t)$ on the initial state.
Moreover, the normalization $\normalization(t)$ is approximately constant, a feature relevant for sampling later on.


\emph{Quantum clock (I)}---
While the above considerations are completely general, a practical implementation relies on a suitable choice of the clock.
It is convenient to consider the harmonic oscillator~\cite{Coppo2024}
\begin{align}
	\hamiltonianclock 
	&= \Omega\sum_{n=0}^{\numcl-1} 2^{n} \frac{( 1-\sigzcl_n )}{2} 
    = \Omega \sum_{Z=0}^{\dimclock-1} Z \dyadclock{Z}{Z},
    \label{eq:clock_hamiltonian_qubits}
\end{align}
regularized to a $\dimclock=2^{\numcl}$ dimensional Hilbert space.
It is diagonal in the computational basis $\ketclock{Z}$, where the integer $Z$ is identified
with the bit string encoding its binary representation.
The energy scale $\Omega$ induces periodic dynamics of the clock with period $T_0 = 2\pi / \Omega$, which is also imprinted on the relational dynamics.
Consequently, any simulation time $T$ should be shorter than $T_0$ in general.
Due to the logarithmic scaling of $N_C$ with $T$ this can be guaranteed by adding only $\mathcal{O}(1)$ qubits to the clock at fixed $\Omega$.
Similar to the explanatory example, the initial state is chosen as equally weighted superposition of all basis states $\ketclock{\stateclock(0)}=\ketclock{+}^{\otimes \numcl} \propto \sum_{Z}\ketclock{Z}$ whose evolution under the clock Hamiltonian reads
\begin{align}
	\ketclock{\stateclock(t)} = 
    \frac{1}{\sqrt{\dimclock}} 
    \sum_{Z=0}^{\dimclock-1} e^{-it\Omega Z} \ketclock{Z} \, .
\end{align}
Those states obey $\sum_Z\ketclock{\stateclock(t(Z, 0)}\propto \ketclock{0}$.
This choice of the clock motivates an intuitive picture how the dynamics is encoded in $\ketglobal{\stateglobal}$. 
Expanding the latter in the clock's energy eigenstates and an arbitrary basis $\{ \ketsystem{\statesystem_n} \}_n$ of $\hilbertsystem$ yields $\ketglobal{\stateglobal} = \sum_{Z,n} c_{Zn} \ketglobal{Z, \statesystem_n}$.
In the corresponding expansion of the relational state $\ketsystem{\statesystem(t)}\propto \braketclockglobal{\stateclock(t)}{\stateglobal} = \sum_nc_n(t)\ketsystem{\statesystem_n}$
(assuming nearly constant normalization) the time-dependent coefficients are given by the finite Fourier series $c_n(t) = \sum_Z c_{Zn} e^{it\Omega Z}$.
In particular, within the above basis, the global Hamiltonian is akin to a Floquet matrix~\cite{Grossmann2018}.


\emph{Engineering interaction (II)}---
For outlining the construction of the interaction, we restrict ourselves to mimicking potentials of the form $\potentialsystemwanted(t) = \envelope(t) \interactionsyspart$ for the sake of presentation. 
In the quasi-eigenstate approximation~\eqref{eq:system_hamiltonian_quasi_eigenstate_approximation}, such a potential can be constructed by setting 
\begin{align}
    \interaction = \interactionclockpart \otimes \interactionsyspart \, .
\end{align}
Here, $\interactionclockpart$ satisfies $\melclock{\stateclock(t)}{\interactionclockpart}{\stateclock(t)} \approx \envelope(t)$ on the domain $[0,T]$ and can be explicitly obtained as
\begin{align}
    \interactionclockpart 
    &= \frac{\dimclock}{T_0} \int_0^{T_0} \dd{t} \enveloperecon(t) 
    \projectorclock(t) \,,
    \label{eq:interaction_clock_part}
\end{align}
where the Fourier coefficients $\enveloperecon_K$ of $\enveloperecon(t) = \sum_{K=-(\dimclock-1)}^{\dimclock-1} \enveloperecon_K e^{-i\Omega K t}$ can be related via $h_K= \frac{\dimclock}{T_0(\dimclock - \abs{K})} \int_0^{T_0} \dd{t} \envelope(t) e^{i \Omega K t}$ for $\abs{k} < \dimclock$. Such a choice minimizes the difference between $\envelope(t)$ and $\melclock{\stateclock(t)}{\interactionclockpart}{\stateclock(t)}$ over the entire interval $[0,T_0]$. Details and relations for smaller time spans (like $[0,T]$) are given in the Supplemental Material.
The operator $\interactionclockpart$ is reminiscent of the Pegg-Barnett phase operator~\cite{Barnett1989, PerezLeija2016} and can be straightforwardly decomposed into Pauli-strings (see Supplemental Material).


\emph{Finding global state (III)}---
Before determining the global state, which is subject to various constraints as detailed below, the question, whether such global states $\ketglobal{\stateglobal}$ exist at all, arises.
In case of a generic system Hamiltonian and despite the simple form of the clock Hamiltonian, however, we expect the presence of the interaction to render the global Hamiltonian to be sufficiently generic as well, such that typicality arguments apply.
Minimizing the energy variance of the global state is equivalent to choosing it from some microcanonical energy shell of the corresponding width.
In the bulk of the spectrum, those shells host exponentially (in $\numcl$) many global energy eigenstates, all of which will be highly entangled between system and clock and hence allow for encoding rich relational dynamics.
Contrarily, the constraint~\eqref{eq:relational_pure_state} for given $\ketsystem{\statesystem}$ and $\ketclock{\stateclock}$ is heavily underdetermined with a space of solutions of codimension $\dim \hilbertsystem$. 
Consequently, for a large enough clock, there will be microcanonical shells, which intersect with the constraint and host enough entangled states to provide a wide range of effective time-dependent potentials $\potentialsystem(t)$.

\begin{figure*}[t] 
    \centering
    \includegraphics[width=\textwidth]{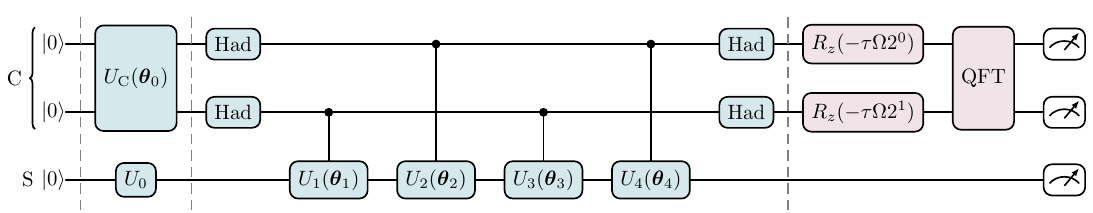}
    \caption{
    Exemplary quantum circuit for the preparation of $\ketglobal{\stateglobal}$ (blue shade) and the projective measurements (red and white shade): starting from an initial $\ketglobal{0}$ state (first dashed line), the state $\ketglobal{\stateglobal_{\txtinit}}$ is prepared (second dashed line), followed by the (variational) preparation of $\ketglobal{\stateglobal}$ (third dashed line) and the subsequent measurements.}
    \label{fig:circuit}
\end{figure*}

The main component for a practical implementation, however, relies on the specific method used for determining an approximate energy eigenstate of $\hamiltonianglobal$. 
Here, we opt for a variational ansatz $\ketglobal{\stateglobal(\prmvqa)} = \unitaryvqa(\prmvqa) \ketglobal{\stateglobal_{\txtinit}}$ and classical optimization~\cite{Cerezo2021} as a VQA for finding an appropriate global state through minimization of a cost function $\costfunc(\prmvqa)$. 
Ensuring the correct initial condition $\ketsystem{\statesystem(0)} = \ketsystem{\statesystemwanted(0)} = \hat{U}_{0} \ketsystem{0}^{\otimes \numsys}$ for $\numsys$ system qubits is facilitated by implementing 
\begin{align}
    \ketglobal{\stateglobal_{\txtinit}} 
    &= \unitaryclockinit(\prmvqa_0) \otimes \hat{U}_{0} \ketglobal{0} \\
    &\propto \left( \sum_{Z=0}^{\dimclock-1} d_Z(\prmvqa_0) \ketclock{\stateclock(t(Z,0))} \right) \otimes \ketsystem{\statesystemwanted(0)}
\end{align}
before the variational procedure. 
Here, $\unitaryclockinit$ is used to guarantee approximately equal probabilities $\abs{d_Z(\prmvqa_0)}^2 \approx \normalization(t(Z,0)) \approx \mathrm{const}$.  
Restricting the variational manifold to global states with the correct initial system state corresponds to circuits fulfilling
\begin{align}
    \left[ \projectorclock(0), \unitaryvqa(\prmvqa) \right] 
    &\propto \left[ \dyadclock{+}{+}^{\otimes \numcl}, \unitaryvqa(\prmvqa) \right] 
    = 0 
\end{align}
for all $\prmvqa$. 
Thus, we employ controlled-unitaries, which are applied for clock qubits being in the state $\ketclock{-_n}$. Two Hadamard gates (Had)~\cite{Nielsen2012} on the control qubit transform a usual control in the computational basis to the desired one.
An exemplary quantum circuit, showing some similarity with quantum phase estimation~\cite{Nielsen2012}, for the variational part is given in Fig.~\ref{fig:circuit}. 
Classical optimization algorithms are used to minimize the cost function
\begin{align}
    \costfunc(\prmvqa) 
    &= \variance_{\stateglobal(\prmvqa)}(\hamiltonianglobal) + \weight \melglobal{\stateglobal(\prmvqa)}{\errorop}{\stateglobal(\prmvqa)} 
    \,,
    \label{eq:cost_function}
\end{align}
while quantum devices can be used to obtain expectation values. 
A freely tunable weight $\weight \geq 0$ can be used to further minimize errors due to the approximation~\eqref{eq:system_hamiltonian_quasi_eigenstate_approximation} at the expanse of increasing the energy variance. 
For a fixed simulation time $T$, 
Eq.~\eqref{eq:time_energy_uncertainty} provides a natural way to set a target threshold for the energy variance and to empirically find convenient values for $\weight$.
The optimal parameter set $\prmvqa^*$ is determined through the minimum $\costfunc(\prmvqa^*) = \text{min}$. 
Once optimization is completed, the global state can be repeatedly prepared for the extraction of dynamical information.


\emph{Sampling dynamics (IV)}---While the relational state~\eqref{eq:relational_pure_state} is defined for arbitrary times $t$, only a finite number of discrete times can be measured at fixed $\tau$ in an actual experiment which implements projective measurements. To be specific, as the set $\{ \ketclock{\stateclock(t(Z,\tau)} \}$ with $\Delta t = T_0/\dimclock$ provides a complete basis for $\hilbertclock$, the global state can be expressed as 
\begin{align}
    \ketglobal{\stateglobal}
    &= \sum_Z \sqrt{\normalization(t(Z,\tau))} \ketclock{\stateclock(t(Z,\tau))} \otimes \ketsystem{\statesystem(t(Z,\tau))} \,.
\end{align}
Furthermore, these clock states can be transformed back to the computational basis through $\unitarysampling(\tau) \ketclock{\stateclock(t(Z,\tau))} = \ketclock{Z}$ by means of the quantum Fourier transform (QFT) in $\unitarysampling(\tau) = \qft \, e^{i \tau \hamiltonianclock}$.
Consequently, after applying $\unitarysampling(\tau)$ on $\ketglobal{\stateglobal}$, we can stochastically sample the relational system states $\ketsystem{\statesystem(t(Z,\tau))}$ (see Fig.~\ref{fig:circuit}). 
Although the computational basis is finite, any point in time can be sampled by simply advancing the clock by an amount $\tau$. Such a transformation corresponds to single qubit rotations for the given clock Hamiltonian~\eqref{eq:clock_hamiltonian_qubits}. 
For an optimal sampling of all relevant times, the measurement probability of different times should be approximately equal, namely $\sqrt{\normalization(t)} \approx \text{const}$. 
Any change of the normalization depends explicitly on the accuracy of approximation~\eqref{eq:system_hamiltonian_quasi_eigenstate_approximation} as the effective system Hamiltonian, because the rate of change reads $\dv*{\normalization(t)}{t} = -i \melglobal{\stateglobal}{ [ \interaction, \projectorclock(t) ] }{\stateglobal}$. 
As discussed above, our construction Eq.~\eqref{eq:interaction_clock_part} provides a good generic candidate for large clock dimension $\dimclock$ to reduce such changes. 
For a vanishing interaction $\interaction=0$, all relational times have the same sampling probability.

\begin{figure}
    \centering
    \includegraphics[width=1\linewidth]{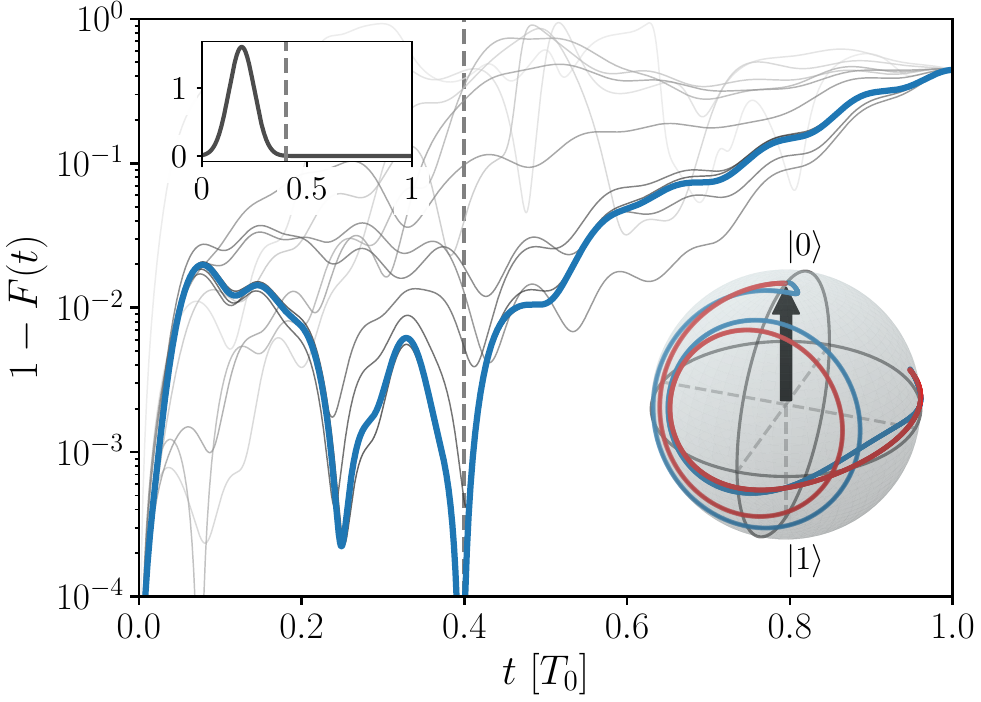}
    \caption{
    Infidelity $1-\fidelity(t)$ of relational qubit dynamics to exact evolution for a time-dependent pulse of the form $g(t) = c \, e^{-(t-t_0)^2/2s^2}$ (inset). Here, we choose $\omega_0=1$, $c=1.6$, $t_0=1.14\pi$, $s=0.36\pi$ and $\numcl=4$, $\Omega=1/3$, $T_0=6\pi$ for the clock.
    Vertical dotted lines denote the desired simulation time $T = 0.4 T_0$. 
    The variational progression (increasingly darker gray lines) with $\weight=0.02$ towards the (at least locally) optimal solution $\ketglobal{\stateglobal}$ yields a 
    final standard deviation of energy  $\sqrt{\variance_{\stateglobal}(\hamiltonianglobal)} \approx 0.062$, amounting to $T^2 \variance_{\stateglobal}(\hamiltonianglobal) \approx 0.22$ in Eq.~\eqref{eq:time_energy_uncertainty}. 
    The evolution of $\ketsystem{\statesystemwanted(t)}$ (red) and $\ketsystem{\statesystem(t)}$ (blue) are also shown on the Bloch sphere for times up to $T$. Details of the quantum circuit can be found in the Supplemental Material.
    }
    \label{fig:infidelity_example}
\end{figure}

\emph{Example}---
Although abundant (approximate) spectral degeneracies are expected to be only present in many-body settings for generic Hamiltonians, we validate our conceptual approach by a numerical demonstration even for a single driven qubit with $\hamiltoniansystem = \omega_0 \sigz/2$. Initialized in $\ketsystem{\statesystemwanted(0)} = \ketsystem{0}$, a Gaussian pulse $g(t)$ induces population transfer through $\potentialsystemwanted(t) = g(t) \sigx$.
The infidelity of the relational state to the exact dynamics for a fixed set of parameters is shown in Fig.~\ref{fig:infidelity_example}. 
While relational dynamics with weak couplings can be treated perturbatively~\cite{Gemsheim2024}, we employ a strong drive outside of this regime.
The true final population on both levels is approximately equal for times $t \geq T$, in contrast to the periodic relational dynamics. As a result, the infidelity gradually increases towards $t \to T_0$. Within the desired time frame $t\in[0,T]$, however, we obtain a fidelity $F(t)$ of $>98\,\%$.

\emph{Conclusion}---
The above example provides a successful proof-of-principle for the algorithm based on relational dynamics and sampling of time steps via projective measurements proposed in this letter.
It allows for simulating time-dependent Hamiltonians at arbitrary times without the need for increasing resources in terms of additional clock qubits.
While the latter nevertheless scales similarly as conceptually related algorithms, 
our approach only requires the preparation of a global eigenstate at arbitrary energy in contrast to a ground state, but at the cost of additional constraints.
Even though a quantitative analysis of the complexity of the presented algorithm is beyond the scope of this letter, we expect the preparation of the global state to be (at least) as hard as the ground state preparation of other approaches.
Nevertheless, our approach provides an alternative way of simulating quantum dynamics, which potentially could be more resilient to noise and decoherence.
This becomes particularly relevant when studying late-time behavior. By 
shifting the weight in the global state to terms corresponding to large times, via imaginary-time propagation for example~\cite{McArdle2019, Motta2019, YeterAydeniz2020, Lin2021, Sun2021, Jouzdani2022, Gluza2024}, it might be possible to access them more efficiently.

The main bottleneck for a successful practical implementation for large systems is the variational determination of the global eigenstate~\cite{Kungurtsev2024, Catli2025}. 
This was demonstrated here for a minimal example only, but we expect more elaborate approaches for state preparation to allow for treating generic systems.
For instance, this could be achieved by tailoring the ansatz for the variational principle to the system at hand, by using machine learning techniques~\cite{Sajjan2021, Du2022} or by employing filtering techniques for preparation of "microcanonical" states~\cite{Sierant2020, Garratt2024}. 
Ultimately, we note that our conceptual approach is not necessarily bound to qubit- or gate-based architectures and could, in principle, be implemented on other platforms as well.
\\

\emph{Acknowledgments}---
S.G.~thanks J.~M.~Rost for insightful discussions and collaborations in related projects. 
F.F.~acknowledges support from the European Union's Horizon Europe program under the Marie Sk{\l}odowska Curie Action GETQuantum (Grant No. 101146632).
\\


\bibliography{quantum_dynamics_sampler}

\pagebreak
\clearpage
\onecolumngrid

\setcounter{equation}{0}
\setcounter{figure}{0}
\setcounter{table}{0}
\setcounter{page}{1}
\makeatletter
\renewcommand{\theequation}{S\arabic{equation}}
\renewcommand{\thefigure}{S\arabic{figure}}
\renewcommand{\bibnumfmt}[1]{[S#1]}
\renewcommand{\citenumfont}[1]{S#1}

\newcommand{\raiseop}{\hat{R}}
\newcommand{\lowerop}{\hat{L}}


\section{Supplemental Material}

\subsection{Correction term to quasi-eigenstate approximation}

If the clock state (approximately) commutes with the interaction, $[\projectorclock(t), \interaction]\approx 0$ for all relevant times, then 
\begin{equation}
    \potentialsystem(t) \ketsystem{\statesystem(t)}
    \approx \melclock{\stateclock(t)}{\interaction}{\stateclock(t)}  \ketsystem{\statesystem(t)} \,.
\end{equation}
It will not be possible to always fulfill this condition exactly, but it provides a great starting point for a constructive search for interaction engineering.
The splitting
\begin{align}
    \melclockglobal{\stateclock(t)}{\interaction}{\stateglobal} 
    &= \melclock{\stateclock(t)}{\interaction}{\stateclock(t)} \braketclockglobal{\stateclock(t)}{\stateglobal}  + \melclockglobal{\stateclock(t)}{[\projectorclock(t), \interaction]}{\stateglobal} \\
    &= \melclock{\stateclock(t)}{\interaction}{\stateclock(t)} \braketclockglobal{\stateclock(t)}{\stateglobal} + \melclockglobal{\stateclock(t)}{\interaction\projectorclockcomp(t)}{\stateglobal} 
\end{align}
with $\projectorclock + \projectorclockcomp = \identclock$ helps to investigate the remainder. 
If the unwanted term is small, then the normalization can be approximated by $\normalization(t)\approx\normalization(0)$ and we get
\begin{align}
    \frac{\melclockglobal{\stateclock(t)}{\interaction}{\stateglobal}}{\sqrt{\normalization(0)}} 
    &\approx \melclock{\stateclock(t)}{\interaction}{\stateclock(t)} \ketsystem{\statesystem} + \frac{\melclockglobal{\stateclock(t)}{\interaction\projectorclockcomp(t)}{\stateglobal} }{\sqrt{\normalization(0)}} \,.
\end{align}
In order to obtain a linear constraint, we use the term
\begin{align}
    \int_0^T \frac{\dd{t}}{T} \melglobal{\stateglobal}{\projectorclockcomp(t) \interaction \projectorclock(t) \interaction \projectorclockcomp(t)}{\stateglobal} 
    &= \braglobal{\stateglobal} \Biggl[ \interaction \overline{\projectorclock}^{T} \interaction + \overline{ \projectorclock \interaction  \projectorclock \interaction \projectorclock }^{T}  - \interaction \overline{ \projectorclock \interaction \projectorclock }^{T} 
    - \overline{ \projectorclock \interaction  \projectorclock }^{T} \interaction \Biggr] \ketglobal{\stateglobal} \\
    &= \braglobal{\stateglobal} \Biggl[ \interaction \overline{\projectorclock}^{T} \interaction + \overline{ \projectorclock \otimes \melclock{\stateclock}{\interaction}{\stateclock}^2  }^{T} - \Bigl\{ \interaction, \overline{ \projectorclock \otimes \melclock{\stateclock}{\interaction}{\stateclock} }^{T} \Bigr\} 
   \Biggr] \ketglobal{\stateglobal} \\
    &\equiv \melglobal{\stateglobal}{\errorop}{\stateglobal} \, ,
\end{align}
with $\overline{\,\cdot\,}^{T}$ denoting the time average.
Such a term can be easily incorporated into a quantum-classical algorithm. If it is minimized, then we ensure that the length of the correction vector is relatively small over the simulation period $[0,T]$. 
The appearing time averages can be further specified.
For the interaction given in the main text,
\begin{align}
    \errorop
    &= \left[ \interactionclockpart \overline{\projectorclock}^T \interactionclockpart  + \overline{\projectorclock \melclock{\stateclock}{\interactionclockpart}{\stateclock}}^T - \Bigl\{ \interactionclockpart , \overline{\projectorclock \melclock{\stateclock}{\interactionclockpart}{\stateclock}}^T \Bigr\} \right] \otimes \interactionsyspart^2 \,.
\end{align}

\subsection{Time averages}

For the specific clock given in the main text, the time averages appearing in the explicit form of $\errorop$ and the determination of an optimal $\enveloperecon(t)$ can be further specified. 
Expressions like
\begin{align}
    \overline{\projectorclock \cdot \genericfunc}^T
    &= \frac{1}{\dimclock} \sum_{M,N=0}^{\dimclock-1} \dyadclock{M}{N}   \int_0^T \frac{\dd{t}}{T} e^{i\Omega t(N-M)} \genericfunc(t) \\
    &= \frac{1}{\dimclock} \sum_{M,N=0}^{\dimclock-1} \dyadclock{M}{N} \genericfunc_{N-M}  \\
    &= \frac{\genericfunc_0}{\dimclock} \identclock + \frac{1}{\dimclock} \sum_{K=1}^{\dimclock-1} \sum_{B=0}^{\dimclock-1-K} \Biggl[ \dyadclock{B}{B+K} \genericfunc_K + \dyadclock{B+K}{B} \genericfunc_{-K} \Biggr]
\end{align}
depend on the Fourier-like components
\begin{equation}
    \genericfunc_N   
    = \frac{1}{T} \int_0^T \dd{t} e^{i\Omega t N} \genericfunc(t) \,.
\end{equation}
for generic functions $\genericfunc(t)$.
For real-valued functions $\genericfunc(t)$, $\genericfunc_{-N} = \genericfunc^*_N$. From the last line of the time average expression one reads off the Toeplitz property. For such matrices, all elements in off-diagonals have the same value. Thus, a $\mathds{C}^{n \times n}$ Toeplitz matrix is characterized by just $1+2(n-1) = 2n-1$ complex numbers. Since we only consider real-valued functions $\genericfunc(t)$, only Hermitian Toeplitz matrices appear and the number of independent real elements reduces to $2n-1$. These elements are determined through the $\{\genericfunc_K\}$. Decompositions of Toeplitz matrices into Pauli strings is given further below.
\\

To provide an example, the clock state average reads
\begin{align}
    \overline{\projectorclock}^{T} 
    &= \frac{1}{\dimclock} \sum_{M,N=0}^{\dimclock-1} \dyadclock{M}{N} Q_T(N-M) \,.
\end{align}
and features the integral
\begin{align}
    Q_{T}(a) 
    &= \frac{1}{T} \int_0^{T} \dd{t} e^{ia\Omega t} 
    = \frac{i}{a\Omega T} \left( 1 - e^{ia\Omega T} \right) 
    = \frac{1}{ia\Omega T} \left( e^{ia\Omega T} - 1 \right) \\
    &\overset{\Omega=2\pi/T_0}{=} e^{i\pi a T /T_0} \sinc\left(\frac{aT}{T_0}\right) 
\end{align}
with $\sinc(x) = \sin(\pi x)/(\pi x)$ and $Q_T(0) = 1$. A special case consists of $Q_{T_0}(a) = \delta_{a0}$ for integers $a \in \mathds{Z}$. In general, the integral's amplitude decreases as $\sim 1/a$. 

\subsection{Clock part of interaction}

\subsubsection{Operator form and mean values}

From the above calculation for the time averages, we can straightforwardly give 
\begin{align}
    \interactionclockpart 
    &= \dimclock \overline{\projectorclock \enveloperecon}^{T_0} 
    = \sum_{M,N=0}^{\dimclock-1} \dyadclock{M}{N} \enveloperecon_{N-M}
\end{align}
and its mean value 
\begin{align}
    \melclock{\stateclock(t)}{\interactionclockpart}{\stateclock(t)} 
    &= \frac{1}{\dimclock} \sum_{M,N=0}^{\dimclock-1} h_{N-M} e^{-i\Omega t (N-M)} 
    = \sum_{K=-(\dimclock-1)}^{\dimclock-1} \frac{\dimclock - \abs{K}}{\dimclock}  h_{K} e^{-i\Omega K t} \\
    &=: \sum_{K=-(\dimclock-1)}^{\dimclock-1} h'_{K} e^{-i\Omega K t} 
    \,.
\end{align}
The intrinsic periodicity $T_0$ manifests itself in the discrete frequencies in the last expression.

\subsubsection{Best match of effective and given potential}

So far, the choice for $\enveloperecon(t)$ is undetermined. We require the above mean values to match the desired/given $\envelope(t)$ as best as possible for $t \in [0,T]$. To this end, we formulate the variational expression
\begin{align}
    0 
    &\overset{!}{=} \frac{1}{2T} \pdv{\enveloperecon'_M} \int_0^T \dd{t} \Bigl[ \melclock{\stateclock(t)}{\interactionclockpart}{\stateclock(t)} - \envelope(t) \Bigr]^2 \\
    &= \frac{1}{2} \pdv{\enveloperecon'_M} \sum_{K,L=-(\dimclock-1)}^{\dimclock-1} \enveloperecon'_{K} \enveloperecon'_L \underbrace{ \frac{1}{T} \int_0^T \dd{t}  e^{-i\Omega (K+L) t} }_{ = Q_T(-K-L)  } 
     - \pdv{\enveloperecon'_M} \sum_{K=-(\dimclock-1)}^{\dimclock-1} \enveloperecon'_{K} \underbrace{ \frac{1}{T} \int_0^T \dd{t} e^{-i\Omega K t} g(t) }_{ = \envelope_{-K} } \\
    &= \sum_{K=-(\dimclock-1)}^{\dimclock-1} \enveloperecon'_{K} Q_T(-K-M) - \envelope_{-M} \,.
\end{align}
As a result, the coefficients $\enveloperecon'_K$ have to fulfill the linear system of equations
\begin{align}
    \sum_{K=-(\dimclock-1)}^{\dimclock-1}  Q_T(M-K) \enveloperecon'_{K} = \envelope_{M} \,.
\end{align}
Using vectors $\vec{\envelope}$ and $\vec{\enveloperecon'}$ representing the sets $\{{\envelope_K}\}$ and $\{{\enveloperecon'_K}\}$, respectively, we get
\begin{align}
    \mathcal{Q} \vec{\enveloperecon'} = \vec{\envelope} \,,
\end{align}
which yields the solution
\begin{align}
    \vec{\enveloperecon'} = \mathcal{Q}^{-1} \vec{\envelope} \,.
\end{align}
Here, $(\mathcal{Q})_{K,M} = Q_T(M-K)$. For $T=T_0$, we obtain $\enveloperecon_K = \dimclock \envelope_K / (\dimclock - \abs{K})$ for $\abs{K} < \dimclock$.
For any non-vanishing vector $\vec{\gamma} = \sum_{L=-(\dimclock-1)}^{\dimclock-1} \gamma_L \vec{e}_L$ with $\gamma_L \in \mathds{C}$, 
\begin{align}
    \vec{\gamma}^\dag \mathcal{Q} \vec{\gamma} 
    &= \sum_{K,M} \gamma^*_K Q_T(M-K) \gamma_M 
    = \frac{1}{T} \sum_{K,M} \gamma^*_K \gamma_M \int_0^T \dd{t} e^{i\Omega T (M-K)} \\
    &= \frac{1}{T} \int_0^T \dd{t} \abs{ \gamma_M e^{i\Omega t M} }^2 > 0
\end{align}
is always strictly positive for $T>0$, which implies invertibility of $\mathcal{Q}$.

\subsubsection{Terms in $\errorop$}

The operator $\errorop$ features the terms
\begin{align}
    \overline{\projectorclock\cdot \melclock{\stateclock(t)}{\interactionclockpart}{\stateclock(t)} }^T
    &= \frac{1}{\dimclock} \sum_{M,N=0}^{\dimclock-1} \dyadclock{M}{N}   \int_0^T \frac{\dd{t}}{T} e^{i\Omega t(N-M)} \sum_{K=-(\dimclock-1)}^{\dimclock-1} \enveloperecon'_{K} e^{-i\Omega K t} \\
    &= \frac{1}{\dimclock} \sum_{M,N=0}^{\dimclock-1} \dyadclock{M}{N} \sum_{K=-(\dimclock-1)}^{\dimclock-1} \enveloperecon'_{K} Q_T(N-M-K)
\end{align}
and
\begin{align}
    \overline{\projectorclock\cdot \melclock{\stateclock(t)}{\interactionclockpart}{\stateclock(t)}^2 }^T
    &= \frac{1}{\dimclock} \sum_{M,N=0}^{\dimclock-1} \dyadclock{M}{N}   \int_0^T \frac{\dd{t}}{T} e^{i\Omega t(N-M)} \sum_{K,L=-(\dimclock-1)}^{\dimclock-1} \enveloperecon'_{K} \enveloperecon'_{L} e^{-i\Omega (K+L) t} \\
    &= \frac{1}{\dimclock} \sum_{M,N=0}^{\dimclock-1} \dyadclock{M}{N} \sum_{K,L=-(\dimclock-1)}^{\dimclock-1} \enveloperecon'_{K} \enveloperecon'_{L} Q_T(N-M-K-L) \,,
\end{align}
which possess the Toeplitz property as well.


\subsection{Pauli decomposition of Toeplitz matrices} 

In order to implement the Toeplitz matrices and multiplications with other operators, we need a way to decompose these matrices in terms of Pauli operators
\begin{align}
	\ident &= \begin{pmatrix}
		1 & 0 \\
		0 & 1
	\end{pmatrix} \\
	\raiseop &= \begin{pmatrix}
		0 & 1 \\
		0 & 0
	\end{pmatrix} = \frac{1}{2} (\sigx + i\sigy) = \lowerop^\dag \\
	\lowerop &= \begin{pmatrix}
		0 & 0 \\
		1 & 0
	\end{pmatrix} = \frac{1}{2} (\sigx - i\sigy) = \raiseop^\dag \\
	\sigx &= \begin{pmatrix}
		0 & 1 \\
		1 & 0
	\end{pmatrix} = \raiseop + \lowerop \\
	\sigy &= \begin{pmatrix}
		0 & -i \\
		i & 0
	\end{pmatrix} = -i\raiseop + i\lowerop = -i (\raiseop - \lowerop) \\
	\sigz &= \begin{pmatrix}
		1 & 0 \\
		0 & -1
	\end{pmatrix} \,.
\end{align}
We consider a matrix $A_{n,k}$ with dimension $2^n \times 2^n$ in the computational basis, ones on the $k$th off-diagonal and zeros on all other entries. This is a Toeplitz matrix and $k$ can take values from $[-2^n-1,2^n-1]$. If $\abs{k} > 2^n$, then $A_{n,k}$ has only vanishing elements. The identity matrix is obtained from
\begin{align}
	A_{n,0} = \ident_{2^n} = \ident^{\bigotimes n} \,.
\end{align}
For $k \neq 0$, we use the binary representation 
\begin{align}
	k = s \sum_{m=0}^{n-1} 2^m k_m 
\end{align}
with $s$ being the sign bit. 
If the most significant bit equals one ($k_{n-1}=1$, implying $\abs{k}\geq 2^{n-1}$ and non-zero matrix elements only in the off-diagonal quadrants of the matrix), we find
\begin{align}
	A_{n, k>0} 
	&= \raiseop \otimes A_{n-1, k - 2^{n-1}} \,,\\
	A_{n, k<0} 
	&= \lowerop \otimes A_{n-1, k + 2^{n-1} } \,.
\end{align} 
If $k_{n-1}=0$ (implying $\abs{k} < 2^{n-1}$), then 
\begin{align}
	A_{n, k>0} 
	&= \ident \otimes A_{n-1, k} + \raiseop \otimes A_{n-1, k - 2^{n-1}} \,,\\
	A_{n, k<0} 
	&= \ident \otimes A_{n-1, k} + \lowerop \otimes A_{n-1, k + 2^{n-1}} \,.
\end{align}
With
\begin{align}
	A_{1,0} = \ident \qquad\quad A_{1,1} = \raiseop \qquad\quad A_{1,-1} = \lowerop \,,
\end{align}
each off these matrices can be recursively determined as tensor products from Pauli operators. Using the sign bit allows us to express everything with the compact formula
\begin{align}
	A_{n, k} 
	&= \delta_{k_{n-1},0} \ident \otimes A_{n-1, k} + \frac{(1 - \delta_{k,0})}{2} (\sigx + i s \sigy) \otimes A_{n-1, k - s2^{n-1}}
\end{align}
with $A_{0,k} = 1$. 
The same recursive structure can be used to determine the Pauli strings representing operators with Toeplitz form in the computational basis, such as $\interactionclockpart$ in the main text. 
For hermitian and anti-Hermitian contributions, we use
\begin{align}
	B_{\pm,n,k} 
	&= \left(  A_{n, k} \pm A_{n, -k} \right) \\
	&= \delta_{k_{n-1},0} \ident \otimes B_{\pm,n-1,k}  \nonumber\\
	&\quad + \frac{(1 - \delta_{k,0})}{2} \Biggl[ \sigx \otimes B_{\pm,n-1,k-s2^{n-1}} + is \sigy \otimes B_{\mp,n-1,k-s2^{n-1}} \Biggr]
\end{align}
with $B_{\pm,0,k\neq 0} = 1 \pm 1$. 
Diagonals need to be treated with a factor of $1/2$, because $B_{+,n,0} = 2 \ident$.


\subsection{Variational circuit for example in main text}

The quantum circuit used for the numerical demonstration resembles the one in Fig.~\ref{fig:infidelity_example} of the main text for four clock qubits ($\numcl=4$). Its main part $\unitaryvqa$ comprises five layers ($n_{\text{reps,S}}=5$) of controlled unitaries (each with three parametrized angles) acting on the system qubit, where each clock qubit is taken as individual control. Furthermore, $\unitaryclockinit$ consists of two layers ($n_{\text{reps,C}}=2$) of controlled $R_z$-rotations between neighboring clock qubits with periodic boundary conditions (see below), which are sandwiched between single qubit $R_y$-rotations.
We use the specific gate definitions
\begin{align}
    R^{(i)}_y(\vartheta) &= e^{-i(\vartheta/2)\sigy_i} \\
    R^{(i)}_z(\vartheta) &= e^{-i(\vartheta/2)\sigz_i} \\
    U^{(i)}(\boldsymbol{\vartheta}) &= \begin{pmatrix}
        \cos(\vartheta_1/2) & -e^{i\vartheta_2} \sin(\vartheta_1/2) \\
        e^{i\vartheta_3} \cos(\vartheta_1/2) & e^{i(\vartheta_2+\vartheta_3)} \sin(\vartheta_1/2) \\
    \end{pmatrix}^{(i)} \\
    \text{c}_iA^{(j)} &= \dyad{0}_i \otimes \ident_2^{(j)} + \dyad{1}_i \otimes A^{(j)} \\
    \text{LHad} &= \bigotimes_{n=0}^{\numcl-1} \text{Had}^{(i)} \\
    \text{L}R_y(\boldsymbol{\vartheta}) &= \bigotimes_{n=0}^{\numcl-1} R^{(i)}_y(\vartheta_i) \\
    \text{Lc}R_z(\boldsymbol{\vartheta}) &= c_{\numcl-1}R^{(0)}_z(\vartheta_i) \prod_{n=0}^{\numcl-2} c_nR^{(n+1)}_z(\vartheta_i) 
\end{align}
in order to construct the variational quantum circuit from
\begin{align}
    U_C(\{\boldsymbol{\nu}_k\},\{\boldsymbol{\xi}_l\})
    &= \text{L}R_y(\boldsymbol{\nu}_0) \prod_{a=1}^{n_{\text{reps,C}}} \left[ \text{Lc}R_z(\boldsymbol{\xi}_a) \, \text{L}R_y(\boldsymbol{\nu}_a) \right] \\
    \hat{U}_{0} = \ident_2 \\
    \unitaryvqa(\{\prmvqa_{n,a}\}) &= \text{LHad} \, \left( \prod_{a=1}^{n_{\text{reps,S}}} \left[ \prod_{n=0}^{\numcl-1} \text{c}_n  U^{(S)}(\prmvqa_{a,n}) \right]  \right) \, \text{LHad} 
    \,.
\end{align}
In total, 80 rotation angles can be tuned within the variational optimization. We note that this is not an optimal circuit structure nor does it yield a better performance than simple Trotter evolution for the system qubit in terms of resources. Our intention is a proof-of-principle to generally show the encoding of the system dynamics in a global state through our proposed method. 


\end{document}